\documentclass[]{raa} 

\usepackage{aas_macros}
\usepackage{graphicx,times}
\usepackage{natbib}
\usepackage{amssymb,amsmath}
\bibpunct{(}{)}{;}{a}{}{,}

\usepackage{hyperref}
\hypersetup{colorlinks = true, linkcolor = green, anchorcolor = red, citecolor = blue, filecolor = red, pagecolor = red, urlcolor = red}

\newcommand{\cpm}{{\small CUBEP$^3$M}}
\def\bq{\begin{equation}}
\def\eq{\end{equation}}
\newcommand{\bqa}{\begin{eqnarray}}
\newcommand{\eqa}{\end{eqnarray}}

\def\smwidth{0.48\textwidth}
\def\smhwidth{0.48\textwidth}

\begin{document}


\title{Cosmological neutrino simulations at extreme scale}

\author{
J.D. Emberson\inst{1,2,3},
Hao-Ran Yu\inst{4,1,5},
Derek Inman\inst{1,6},
Tong-Jie Zhang\inst{5,7},
Ue-Li Pen\inst{1,8,9,10},
Joachim Harnois-D\'{e}raps\inst{11,12},
Shuo Yuan\inst{13},
Huan-Yu Teng\inst{5},
Hong-Ming Zhu\inst{14},
Xuelei Chen\inst{14}, and
Zhi-Zhong Xing\inst{15,16}
}

\institute{ 
Canadian Institute for Theoretical Astrophysics, University of Toronto, M5S 3H8, Ontario, Canada; {\it emberson@astro.utoronto.ca}\\ \and
Department of Astronomy \& Astrophysics, University of Toronto, Toronto, Ontario M5S 3H4, Canada\\ \and
ALCF Division, Argonne National Laboratory, Lemont, IL 60439, USA\\ \and
Kavli Institute for Astronomy \& Astrophysics, Peking University, Beijing 100871, China\\ \and
Department of Astronomy, Beijing Normal University, Beijing 100875, China; {\it tjzhang@bnu.edu.cn}\\ \and
Department of Physics, University of Toronto, Toronto, Ontario M5S 1A7, Canada\\ \and
School of Physics and Electric Information, Shandong Provincial Key Laboratory of Biophysics, Dezhou University, Dezhou 253023, China\\ \and
Dunlap Institute for Astronomy and Astrophysics, University of Toronto, Toronto, ON M5S 3H4, Canada\\ \and
Canadian Institute for Advanced Research, Program in Cosmology and Gravitation\\ \and
Perimeter Institute for Theoretical Physics, Waterloo, ON, N2L 2Y5, Canada\\ \and
Department of Physics \& Astronomy, University of British Columbia, Vancouver, BC V6T 1Z1, Canada\\ \and
Scottish University Physics Alliance, Institute for Astronomy, University of Edinburgh, EH9 3HJ, Scotland, UK\\ \and
Department of Astronomy, Peking University, Beijing 100871, China\\ \and
Key Laboratory for Computational Astrophysics, National Astronomical Observatories, Chinese Academy of Sciences, Beijing 100012, China\\ \and
School of Physical Sciences, University of Chinese Academy of Sciences, Beijing 100049, China\\ \and
Institute of High Energy Physics, Chinese Academy of Sciences, Beijing 100049, China\\
}

\abstract{
Constraining neutrino mass remains an elusive challenge in modern physics. 
Precision measurements are expected from 
several upcoming cosmological probes of large-scale structure.
Achieving this goal relies on an equal level of precision from theoretical predictions of neutrino clustering. 
Numerical simulations of the non-linear evolution of cold dark matter and neutrinos play 
a pivotal role in this process. We incorporate neutrinos into the cosmological N-body code
\cpm\ and discuss the challenges associated with pushing to the extreme scales
demanded by the neutrino problem. We highlight code optimizations made to exploit
modern high performance computing architectures and present a novel method of data compression 
that reduces the phase-space particle footprint from 24 bytes in single precision to roughly 9 bytes. 
We scale the neutrino problem to the Tianhe-2 supercomputer and provide details of our production run, 
named TianNu, which uses 86\% of the machine (13,824 compute nodes). With a total of 2.97 trillion particles, TianNu
is currently the world's largest cosmological N-body simulation and improves upon previous neutrino simulations 
by two orders of magnitude in scale. We finish with a discussion of the unanticipated computational challenges
that were encountered during the TianNu runtime.
}

\authorrunning{Emberson et al.}
\titlerunning{Cosmological neutrino simulations at extreme scale}

\maketitle


\section{Introduction}

The standard model of particle physics predicts the existence of three neutrino flavours:
electron, muon, and tau. These flavours exist as superpositions of three mass eigenstates,
with the generic prediction of the standard model claiming each eigenstate have identically
zero mass. However, various extensions of the standard model exist for which the mass eigenstates
can be non-zero \citep{lesgourgues/pastor:2006}. In this case, \citet{pontecorvo:1958} showed it is
possible  that flavour is not conserved, allowing neutrinos to oscillate between flavour with time. 
This phenomena was  firmly established by observations of the flux of electron neutrinos from the Sun that 
were roughly three times  smaller than expected based on solar models \citep{ahmad/etal:2002}. 
The resolution is that electron 
neutrinos, the only flavour produced in the Sun, oscillate between muon and tau during their passage 
through the Sun, leading to a suppressed flux of electron neutrinos when they arrive at Earth \citep{wolfenstein:1978,
mikheyev/smirnov:1985}. The existence of neutrino oscillations have also been verified from the flux of electron 
and muon neutrinos produced from cosmic ray collisions in the Earth's atmosphere \citep{fukuda/etal:1998}. 
These atmospheric and solar neutrino oscillation experiments were awarded the 2015 Nobel Prize in physics
for their confirmation of massive neutrinos. 

Constraining the absolute mass hierarchy of neutrinos is a challenging problem in physics. 
Oscillation experiments imply that at least two of the three neutrino eigenstates are massive, with minimum masses of 
roughly 10 and 50 meV \citep{capozzi/etal:2016}. Unfortunately, these experiments are only sensitive to the mass-squared 
splittings between mass eigenstates and cannot be used to infer the hierarchy of individual neutrino masses. 
This leaves many open questions into the nature of these fundamental particles. To this end, particle physicists 
have devised numerous experiments that aim to place constraints on individual neutrino masses including observations of the $\beta$ 
decay of tritium \citep{kraus/etal:2005,KATRIN:2011}, or from the possibility of neutrinoless double-$\beta$ decay 
\citep{agostini/etal:2013,EXO:2014} in the event that neutrinos are Majorana particles.

Cosmologists have also been working hard to constrain the neutrino mass hierarchy. Relic neutrinos produced shortly after
the Big Bang are second only to photons as the most abundant particle in the universe. As such, they have the potential
to impact cosmological phenomena including the cosmic microwave background (CMB)
and large-scale structure (LSS). Currently, the best constraints on neutrino mass come from the Planck CMB satellite, 
with an upper bound on the sum of neutrino mass of $\sum m_\nu < 194$ meV \citep{PLANCK:2015}.
Future LSS experiments such as Euclid \citep{EUCLID:2013} and LSST \citep{LSST:2012} are expected to reduce 
this upper bound to the $\sim 40$ meV level using precision measurements of the matter power
spectrum \citep{costanzi/etal:2013}. Another potential LSS probe involves measuring the dipole asymmetry in 
the matter density field that results from the relative flow between cold dark matter (CDM) and 
neutrinos \citep{zhu/etal:2014a,zhu/etal:2014b,inman/etal:2015}.

In preparation of these upcoming probes, theorists must make predictions for
the effect of massive neutrinos on LSS within the non-linear regime
where analytic calculations of the growth of structure break down. They are thus forced to rely on the use of
cosmological structure formation simulations. Such simulations have a mature history in 
cosmology, with the earliest N-body schemes being implemented in the 1970's
\citep[e.g.,][]{peebles:1970,miyoshi/kihara:1975,white:1976,aarseth/etal:1979}.
The general picture is that of a set of point particles evolving under their mutual gravitational interaction
with some combination of particle-mesh (PM), particle-particle (PP), and tree algorithms used for the force calculation.
Since their first inception, cosmological N-body simulations have been widely adopted and optimized for 
high performance computing (HPC) environments, with the current state-of-the-art simulations now reaching
the trillion-particle scale for the pure CDM case \citep{skillman/etal:2014,habib/etal:2016}.

Cosmological N-body simulations that self-consistently coevolve CDM and neutrino particles have only recently begun to mature in
scale \citep[e.g.,][]{brandbyge/etal:2008,viel/etal:2010,bird/etal:2012,villaescusa-navarro/etal:2013,inman/etal:2015,castorina/etal:2015}.
The main technological challenge in this case
is that neutrinos are thermally hot, having velocity dispersions several orders of magnitude greater than
that of CDM. This thermal motion suppresses the ability of neutrinos to gravitationally clump on small scales
and tends to distribute them more uniformly throughout the simulation volume compared to CDM. 
As a result, a tremendous amount of shot noise exists on small scales and can only be reduced by increasing 
the particle count to large numbers. Any attempt to simulate the non-linear interaction between CDM and neutrinos
in a large cosmological volume must overcome this computational burden. Fortunately, some of this burden
is alleviated when scaled to a large number of parallel tasks since the near homogeneity of neutrinos leads to
a computational load that is significantly more balanced than the pure CDM case. Hence, the cosmological neutrino
problem is perfectly suited for modern HPC.

We focus here on optimizing the cosmological N-body code \cpm\ \citep{harnois-deraps/etal:2013} for the extreme scale
demanded by the neutrino problem. We highlight specific code changes relevant for both
the neutrino case as well as the more general class of cosmological simulations. Our method is applied to Tianhe-2
which, as of submission, ranks second on the Top 500 list of 
supercomputers\footnote{\url{https://www.top500.org/lists/2016/06/}}. 
Our production run uses 13,824 compute nodes of Tianhe-2 ($86\%$ of the machine) to evolve $6912^3$ CDM particles
and $13824^3$ neutrino particles in a cubic volume of width 1200 $h^{-1}$Mpc. We name this simulation TianNu, 
or the ``Neutrino Sky''. With a total of 2.97 trillion particles, TianNu is the largest cosmological N-body simulation 
performed to date and is two orders of magnitude larger than previous cosmological neutrino simulations. 
As TianNu was hitherto the most computationally ambitious simulation performed on Tianhe-2, 
we were given two weeks of dedicated access to scale our problem and debug any potential issues with the machine. 
We discuss here the unforeseen challenges that were uncovered as we pushed to the current limit of scientific computing. 

This paper is organized as follows. In Section 2 we provide an overview of the cosmological N-body problem and
document the optimizations made to \cpm\ to adapt its usage for extreme-scale HPC. These include the 
implementation of an MPI pencil decomposition in the long-range PM force solver, the addition of memory-efficient 
nested OpenMP parallelism in the short-range PM and PP force evaluations, and a novel method of data compression
relevant for cosmological simulations. Section 3 presents a weak scaling analysis on the Tianhe-2 and provides details of our
production run. We finish with a discussion of the various technical challenges that were encountered while utilizing
86\% of the machine. The conclusions of our work are summarized in Section 4.


\section{Numerical Method and Optimizations}
\label{sec:method}

\subsection{Code Overview}
\label{subsec:method-cpm}

We begin with a brief overview of the methodology employed by the cosmological code 
\cpm\ used in this study. The text presented here is meant to provide background information
that will augment the code optimizations described in the proceeding subsections. 
We refer the reader to previous works \citep{harnois-deraps/etal:2013,inman/etal:2015} for a more
thorough analysis of the code structure and technical algorithms relevant to both the pure CDM and
neutrino cases. 

\begin{figure}
\begin{center}
\includegraphics[width=\smwidth]{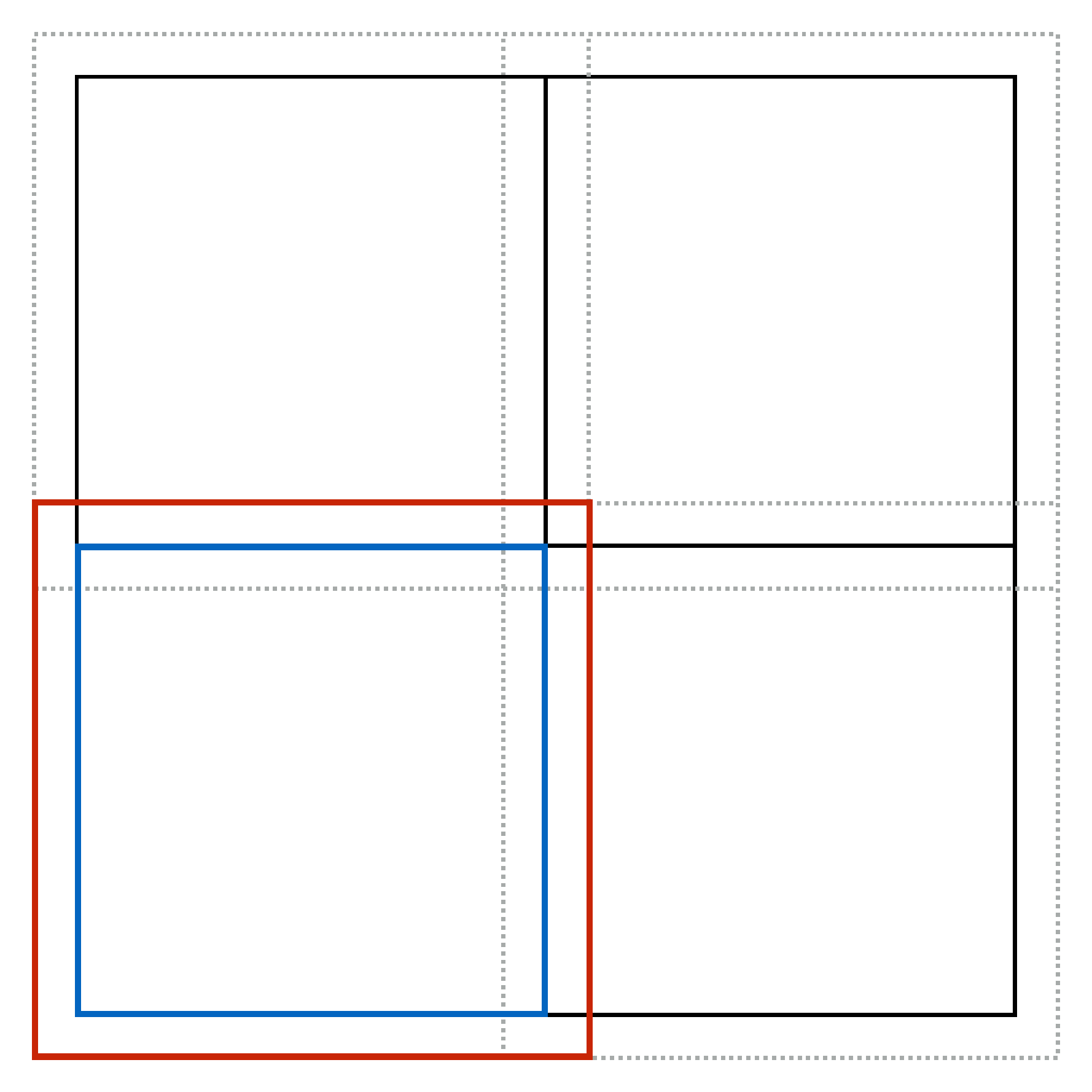}
\caption{Two-dimensional representation of the domain decomposition into tiles within a single node (i.e., MPI task)
of the overall simulation. In this example, we split the node into two tiles per dimension, with boundaries of the
tiles denoted by solid black lines. The extent of the short-range PM force and PP force (see text) are constrained 
within the range of an individual tile so that OpenMP threads can be used to cycle over all tiles on a node. A small 
buffer region (indicated by dotted grey lines) is used to ensure accurate force calculation near tile boundaries.
The solid blue line highlights the bottom left tile with its corresponding buffer highlighted in red. MPI communications
are minimized by requiring their use only during the long-range PM force calculation and at the end of 
each time step when particles are passed between neighbouring nodes.}
\label{fig:tiles}
\end{center}
\end{figure}

Cosmological simulations are parameterized by the physical volume they resolve and the number
of particles they contain. The volume is generally represented as a periodic cube of side length $L$ and
the number of particles expressed as an integer cubed, $N_p = n_p^3$. In the case of \cpm, the domain is 
decomposed into cubes of equal volume, with each cube assigned to a single MPI task. 
With this setup, the number of MPI tasks assigned to the problem is constrained to be a cubed integer, 
$N_{\rm mpi} = n_{\rm mpi}^3$.
We refer to this top level of cubic domain decomposition as {\em nodes} since the usual operation of 
the code assigns one MPI task per compute node. Within each node exists a second level of cubic
subdivision, into equal volume elements called {\em tiles}. Calculations within each node's
volume are done simultaneously over tiles using OpenMP threads, as described in more detail below. 
The user is free to choose the number
of tiles within each node, with the ideal strategy to make this an integer multiple of the number
of threads available to each MPI task. Figure \ref{fig:tiles} shows a two-dimensional representation of the
decomposition into tiles within a single MPI domain. 

The objective of cosmological simulations is to evolve particles from an initial configuration at an
early cosmic time to the present epoch. At each time step, the key quantity to compute is the three-dimensional 
gravitational force field. This is achieved in \cpm\ using both PM and PP methods.
Hybrid codes combining multiple force schemes like this are common in cosmological
applications. Another common choice is to substitute a Tree algorithm in place of the PM method used here. We opt
for a PM method since it is much faster for nearly homogeneous particle distributions,  
as is especially the case with cosmological neutrinos. Additional advantages of the PM method over Tree 
algorithms include reduced memory overhead and ease of parallelizability. The PP force increases 
resolution below the mesh scale where the PM force loses range.

In the PM scheme, the gravitational force is found by solving the Poisson equation:
\bq
\nabla^2 \phi({\bf x}) = 4 \pi G \rho({\bf x}).
\label{eq:poisson}
\eq
Here $\phi$ is the gravitational potential at spatial (comoving) coordinate ${\bf x}$, $\rho$ is the matter density field,
and $G$ is the gravitational constant. The density field is computed by interpolating particles onto a uniform
cubic mesh containing $N_g = n_g^3$ cells. The result is then Fourier transformed and the potential is solved via
\bq
\tilde{\phi}({\bf k}) = - \frac{4\pi G}{k^2} \tilde{\rho}({\bf k}),
\label{eq:poisson-fourier}
\eq
where ${\bf k}$ is a Fourier mode. The gravitational force in Fourier space is related to the potential as
\bq
\tilde{F}({\bf k}) = - i m \tilde{\phi}({\bf k}) {\bf k},
\label{eq:forcek}
\eq
where $m$ is the particle mass and $i$ is the imaginary unit. The three components of the force
in real space are obtained from three inverse Fourier transforms of equation (\ref{eq:forcek}).

\cpm\ splits the PM force computation into two terms: a long-range force term and a short-range force term.
The former is computed on a coarse mesh containing $n_g^3 = (n_p/2)^3$ cells\footnote{In the case of our neutrino
simulations, $n_p$ here refers to the number of neutrino particles which we normally choose to be 8 times greater 
than the number of CDM particles. This is chosen out of the necessity of suppressing the high level of neutrino shot
noise that results from their thermal motion. Both CDM and neutrino particles are interpolated to the same mesh when
computing PM forces.} while the latter uses a mesh that is a factor of 4 times finer in each dimension.
The long-range force is solved over the global simulation volume using Fourier transforms evaluated in parallel over 
all MPI tasks while the latter is solved on the local mesh of each individual tile. This approach minimizes MPI
communication where coarse resolution elements are sufficient in the long-range force calculation while maintaining high resolution
on the small scales that depend only on rank-local shared memory.

The utility of the PM scheme is its speed, with the Fourier transforms being order
$\mathcal{O}(N_g {\rm log}N_g)$. The downside is the force is heavily suppressed below the mesh
scale. In \cpm, this is remedied by appending the short-range PM force with a PP force calculation below the 
grid scale. The PP force is evaluated using a direct pairwise summation:
\bq
F_i = G m_i \sum_{j \neq i} m_j \frac{{\bf x}_j - {\bf x}_i}{| {\bf x}_j - {\bf x}_i|^3},
\label{eq:ppforce}
\eq
where $F_i$ is the force on particle $i$ at spatial location ${\bf x}_i$. The sum is performed over all particles
within the same grid cell as particle $i$ and is of order $\mathcal{O}(\eta_p^2)$ where $\eta_p$
is the typical number of particles within one cell of the short-range interpolation mesh. In order to avoid 
artificial scattering in the N-body problem, the PP force is truncated below some chosen scale called the
softening length, $r_{\rm soft}$.
In \cpm, the softening length is normally chosen to be 0.05 times the average inter-particle spacing; that is,
$r_{\rm soft} = L/(20n_p)$. 

\subsection{MPI Pencil Decomposition}
\label{subsec:method-lrpm}

The first stage in our force evaluation is the long-range PM force. As described above, this is solved 
using MPI Fourier transforms evaluated on the global
interpolation mesh containing $n_g^3 = (n_p/2)^3$ cells. Previously, this was achieved using the FFTW library
\citep{frigo/johnson:2005} with a one-dimensional MPI slab decomposition. This setup divides the global interpolation
mesh into $N_{\rm mpi}$ planes of size $n_g \times n_g \times (n_g/n_{\rm mpi}^3)$. Hence, the problem becomes 
constrained by the requirement that the number of cells per side, $n_g$, be an integer multiple of the
total number of MPI tasks, $n_{\rm mpi}^3$. This constraint scales poorly
since the number of MPI tasks is proportional to the three-dimensional problem size 
and can thus quickly exceed the number of interpolation cells in one dimension. 

To remedy this issue, we have modified \cpm\ so that the long-range parallel Fourier transforms
are handled using the publicly available P3DFFT library \citep{pekurovsky:2012}. P3DFFT employs a 
two-dimensional pencil decomposition where the interpolation mesh is divided into $N_{\rm mpi}$ pencils
of size $n_g \times (n_g/n_{\rm mpi}) \times (n_g / n_{\rm mpi}^2)$. The decomposition constraint is 
significantly alleviated with the requirement that $n_g$ be an integer multiple of only $n_{\rm mpi}^2$ rather 
than $n_{\rm mpi}^3$. An added advantage of the pencil decomposition is that the number of MPI communications
decrease by a factor of $n_{\rm mpi}$ when transforming data between cubic nodes and pencils. 
In particular, a slab (pencil) decomposition requires $n_{\rm mpi}^2$ ($n_{\rm mpi}$) MPI communications 
corresponding to each of the individual nodes a given slab (pencil) intersects in the $xy$ plane. 
The pencil decomposition thus requires larger but fewer MPI communications. 

\subsection{Nested OpenMP Parallelism}
\label{subsec:method-lrpm}

Unlike the long-range PM force, the short-range PM and PP forces are local quantities that can ignore sufficiently
distant parts of the simulation volume. The region within which these forces operate can be completely isolated within
individual tiles. In this case, the short-range PM force is evaluated using an interpolation mesh and Fourier transforms
local to each tile while the PP force is evaluated by looping over each cell in the mesh and identifying its constituent 
particle pairs.  A small buffer region around each tile is used to ensure that the force is accurately computed for
particles near the boundary of the tile (see Figure \ref{fig:tiles}). With this setup, no communication between tiles is
necessary and we perform the loop over tiles using OpenMP threads with dynamic scheduling used to minimize load
imbalance. 

\begin{figure}
\begin{verbatim}
subroutine compute_tile_forces
   call link_list
   !$omp do num_threads(mt)
   do tile = 1, num_tiles
       thread = omp_get_thread_num() + 1
       rho(1:ng, 1:ng, 1:ng, thread) = 0
       call density_interpolation(nt,thread)
       call Fourier_transforms(nt,thread)
       call inverse_force_interpolation(nt,thread)
       call pp_force(nt,thread)
   enddo
   !$omp end do
end subroutine compute_tile_forces
\end{verbatim}
\caption{Pseudocode showing the usage of nested OpenMP threading within each tile to maximize parallel efficiency 
on many-core systems with limited memory. Here ``mt'' represents the number of master threads while ``nt'' is the
number of nested threads available to each master thread. The initialization of the ``rho'' array provides an example of the
memory overhead associated with each master thread. Nested OpenMP loops exist in each of the subroutine calls
with data written only to those arrays initialized by the master thread.  All threaded regions employ dynamic scheduling
to minimize load imbalance. The calculation of the linked list prior to looping over tiles accelerates particle lookup in the 
interpolation and PP stages.}
\label{fig:pseudocode}
\end{figure}

Previously, each tile would be assigned only one thread, meaning the number
of active tiles at any moment equals the number of OpenMP threads. In the usual operation of the code,
this number is equal to the number of cores on a single compute node, or an integer multiple of this if
hyper-threading is possible. Since each active thread must store its local interpolation mesh (as well as other arrays
dedicated to the PP force calculation), this approach 
can become expensive in modern HPC architectures where the strategy is to maximize the number of cores
on a single node. To maximize parallel efficiency with finite memory, we have 
incorporated nested OpenMP parallelism into \cpm\ so that each tile may be handled by multiple threads. 
The memory overhead of the tile is assigned to the master thread while nested threads perform 
the particle interpolations, Fourier transforms, and PP forces in parallel. This change allows us to maximize core
usage, especially at late times in the simulation when the PP force easily dominates the compute time as clustered
objects form.

Figure \ref{fig:pseudocode} provides an overview of the nested OpenMP parallelism within each tile. 
The first stage is the calculation of a linked list
defined on a mesh with the same resolution as the long-range PM interpolation mesh. The linked list is essential in
accelerating particle access in the interpolation stages as well as the PP force evaluation. Next, we use ``mt'' master 
threads to perform the dynamic scheduling over all num\_tile tiles. Each master thread initializes an array, rho, which is
used to store the three-dimensional density interpolation on the local mesh of the tile containing ng cells. Another set of arrays 
pertaining to the PP force evaluation are also initialized by each master thread. The balance between the number of
master and nested threads is dictated by the memory overhead associated with each master thread. 
The short-range PM force calculation is composed of the density interpolation, three Fourier transform calls (one per
dimension), and the inverse force interpolation. Each of these routines is parallelized using ``nt'' nested threads with data written
only to the local arrays initialized by the master thread. More specifically, the density interpolation involves a threaded loop over
all cells in the linked list with a stride in the outer loop used to prevent race conditions that may occur when writing to rho with
a cloud-in-cell (CIC) interpolation. The stride is important since it avoids potentially slow thread locks. 
The inverse force interpolation involves a similar threaded loop over all cells in the linked list. The Fourier transforms are 
evaluated using the threaded versions of the one-dimensional FFTW3 routines. The final stage in each tile is the PP force which again
involves a loop over each cell in the linked list. Each stage shown in Figure \ref{fig:pseudocode} uses dynamic 
scheduling of OpenMP threads to minimize load imbalance.

\subsection{Data Compression}

Our final consideration involves data compression. It is generally important for any application to store snapshots
at various stages during runtime for both checkpointing and data analysis. 
In cosmological simulations, the relevant information usually involves the position and velocity of each particle.
This amounts to 24 bytes per particle for a three-dimensional problem represented in single precision. Simulations now 
reaching the trillion-particle scale thus require tens of TB for a single particle snapshot. This is a formidable 
challenge from the standpoint of both runtime I/O as well as data storage and handling.

We have devised a new method of data compression relevant for cosmological simulations. The first stage
is to construct a linked list based on particle positions in the global simulation volume. 
We use a cubic mesh containing $n_g^3 = (n_p/2)^3$ cells for this purpose. In our case, this step is already 
done prior to the force evaluations (see Figure \ref{fig:pseudocode}) and thus requires
no additional work. Particle information is written for each cell in the linked list in an ordered fashion, as follows. 
First, we write the number of particles in the given cell as an unsigned 1-byte integer\footnote{If the number of 
particles in the cell is equal to or larger than 255, we first write 255 and then write the actual number as a 4-byte
integer.}. Next, for each particle in the cell, we write its three components of position and velocity. 
Positions are compressed by converting them to unsigned 1-byte integers that represent their digitized
offset from the local origin of that cell. Since particles are written in a cell-ordered format this guarantees that 
positions are stored with a precision of $L/(256n_g) = L/(128n_p)$. This truncation is roughly six times finer than the force 
resolution of the simulation specified by $r_{\rm soft}$ so that it should have minimal impact on subsequent
dynamics when used as a restart. Velocities are more difficult to compress due to their unstructured distribution 
within the simulation volume. We choose to compress velocities into 2-byte integers representing their index within
a histogram containing $2^{16}$ equally spaced bins ranging from $-v_{\rm max}$ to $v_{\rm max}$.
Here $v_{\rm max}$ is the maximum (absolute) particle velocity across all ranks and is stored in the header of the output
file.

This process reduces the particle footprint from 24 bytes to roughly 9 bytes. In our method, each MPI task
writes its local volume to separate files in the cell-ordered format described above. In addition, 
different particles species (i.e., CDM, neutrinos) are written to separate files so the species type of each 
particle does not need to be stored. In this case, it is important to compute $v_{\rm max}$ for each species separately 
since they may have vastly different characteristic velocities. For our purposes here, it is sufficient to store only position
and velocity information. More sophisticated compression algorithms would need to be devised for storing additional
quantities (e.g., particle identification tags).

\begin{figure*}
\includegraphics[width=\smhwidth]{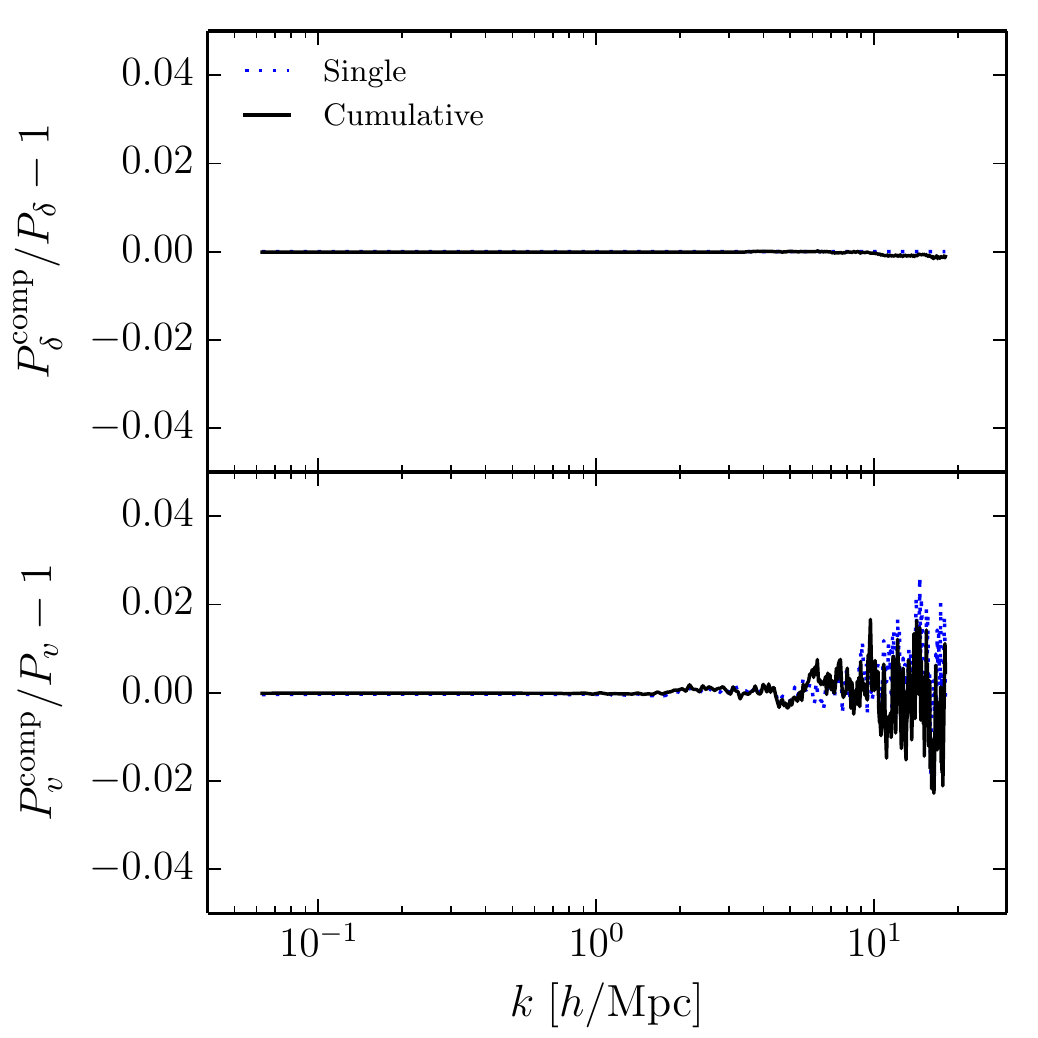}
\includegraphics[width=\smhwidth]{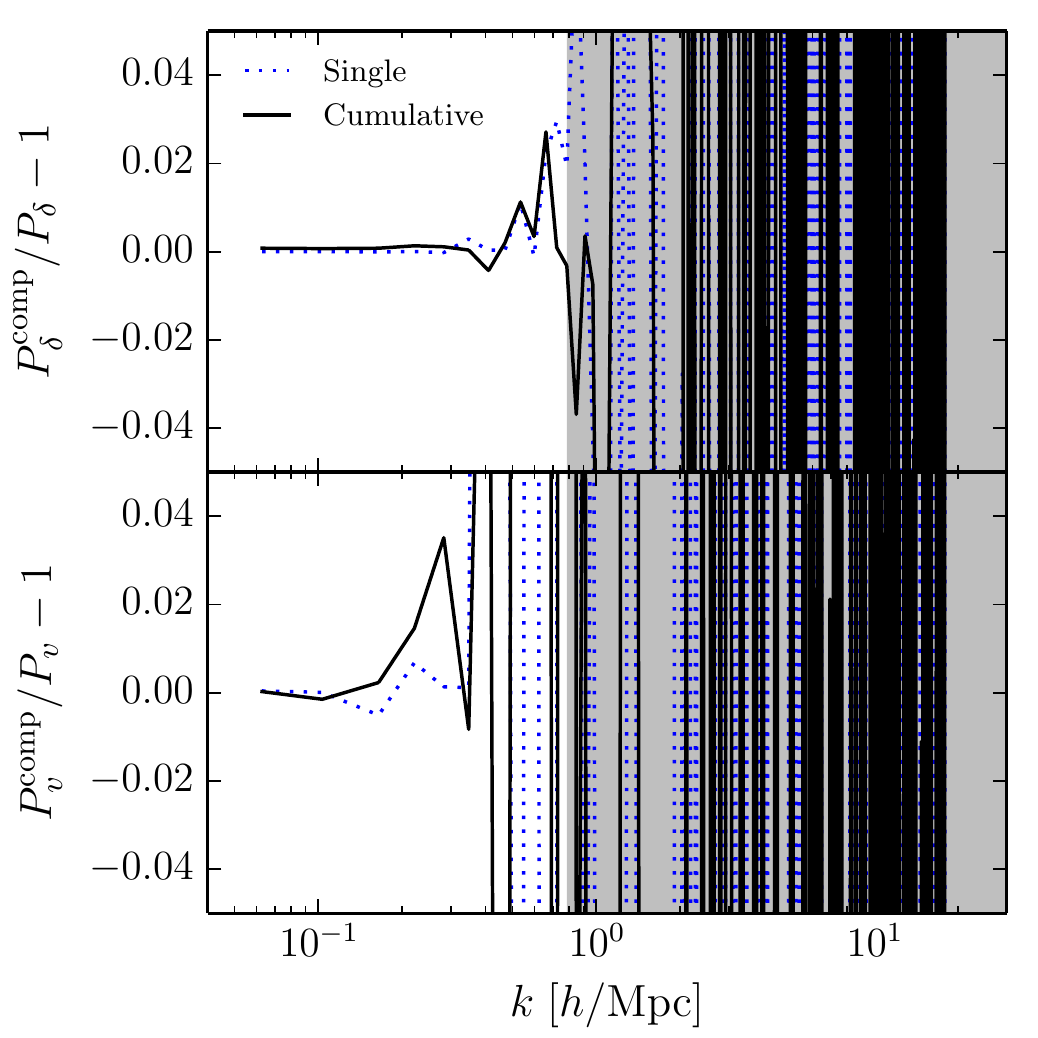}
\caption{Relative difference between density (top panels) and velocity (bottom panels) power spectra between
an uncompressed simulation ($P_{\delta,v}$) and a simulation that restarted from a series of 10 compressed
snapshots ($P^{\rm comp}_{\delta,v}$). Both simulations have identical initial conditions and contain 
$576^3$ CDM plus $1152^3$ neutrino particles in a box of side length 100 $h^{-1}{\rm Mpc}$.
The left plot corresponds to CDM while the right plot corresponds to neutrinos.
Solid black lines denote the relative difference in power  at the final output which includes the cumulative
error from all 10 compressed restarts. For comparison, the dotted blue line shows the relative difference in power that
results from compressing the final output of the uncompressed simulation. The shaded grey regions in the right panels
highlight the scales for which fluctuations in the neutrino measurements are dominated by shot noise rather than 
compression error.}
\label{figure:compress} 
\end{figure*}

We have tested the error associated with a single compression as well as the accumulation in error that results from
restarting multiple times from compressed data. For this purpose, we performed two simulations 
with identical initial conditions pertaining to a cosmological setup with $576^3$ CDM and $1152^3$ 
neutrino particles in a box of side length 100 $h^{-1}{\rm Mpc}$. The first simulation ran to completion 
without any data  compression. The second simulation repeatedly wrote and restarted from compressed data in order to 
gauge the total accumulation in error at the final output. In this case, we tested 10 compressed restarts occurring 
at redshifts $z = \{5, 4.5, 4, 3.5, 3, 2.5, 2, 1.5, 1, 0.5\}$.

Figure \ref{figure:compress} compares  density and velocity power spectra between the two simulations for the final 
output at redshift $z = 0$. The left panels correspond to CDM and show sub-percent agreement in the density 
power spectra for the entire range in $k$ which extends to half the Nyqvist frequency, $k_{\rm Nyq}/2 = 18\ h/{\rm Mpc}$, 
of the $1152^3$ mesh used to sample power. The CDM velocity power is also sub-percent up to about $k_{\rm Nyq}/3$ 
with $2\%$ fluctuations on the smallest scales. For comparison, the dotted blue line shows the relative difference in power
that results from compressing the final snapshot of the uncompressed simulation. Interestingly, the solid black and dotted blue
lines are in good agreement suggesting that the majority of the error is associated with a single compression, rather than
accumulation in error from repeated restarts. 

Neutrino power spectra are shown in the right panels of Figure \ref{figure:compress}. The agreement on the largest scales  
is similar to the CDM case, but large deviations clearly dominate on small scales. However, these deviations are not related
to the compression itself, but are rather attributed to neutrino shot noise. The dashed grey regions highlight the scales for which
$k > k_{\rm noise}$ where $k_{\rm noise}$ is the scale at which the Poisson power, $P_{\rm noise} = 1/\bar{n}$, overtakes
the cosmological density power, $P_\delta$. Here $\bar{n}$ is the mean number density of neutrinos. 
Density and velocity power spectra are intrinsically noisy on scales smaller than $k_{\rm noise}$ meaning 
that deviations seen in the compressed data are not particularly meaningful. Note that CDM does not run into this issue
since the CDM density power is larger than $P_{\rm noise}$ on all scales resolved by the simulation. As discussed before, the
high level of neutrino shot noise is precisely what makes these simulations expensive.

In summary, we find sub-percent agreement in density power spectra for scales $k \lesssim k_{\rm noise}$. Velocity power exhibits
a similar level of agreement, but only up to scales $k \lesssim k_{\rm Nyq}/3$, with percent-level deviations emerging on smaller scales.
The algorithm presented here is well suited for both CDM and neutrinos though other factors such as shot noise must be combated
in the latter case. The agreement between the single-compression case and the test case containing 10 compressed restarts indicates 
that compression has little effect on subsequent measurements of statistical quantities such as density and velocity power spectra. 
While individual particle trajectories are likely to be more discrepant, cosmological observations are statistical in nature and thus afford relatively 
relaxed constraints on compression. This motivates the question of how much precision is required for storing positions and velocity within
simulation memory. We note that a memory-light version of \cpm\ has been developed \citep{yu/pen:2016} that can use adjustable precision 
(8- or 16-bit) for position and velocity, and will address the usage of reduced precision in cosmological applications. 


\section{Results}
\label{sec:results}

\subsection{Weak Scaling}
\label{subsec:weakscaling}

We begin this section with an investigation of the scaling performance of our cosmological neutrino simulations
on Tianhe-2. We pivot our scaling analysis against a single-node simulation that coevolves $288^3$ CDM particles
and $576^3$ neutrino particles in a periodic box of side length $50\ h^{-1}{\rm Mpc}$. We assign a single MPI task
to this node with 24 OpenMP threads used to utilize each of the 24 CPU cores available on the node. Four of these
threads are assigned as master threads that cycle over tiles (see Figure \ref{fig:pseudocode}), with each master thread
spawning six nested threads to handle the inner force computations local to the tile. The MPI domain is divided into $6^3$ tiles
with each assigned a mesh containing $(192+2n_{\rm buf})^3$ cells for computing the short-range PM force and PP force. Here 
$n_{\rm buf} = 24$ is the size of the buffer region used to ensure correct forces near the boundary of the tile.
The long-range PM force is computed on the global $288^3$ mesh. 

\begin{figure}
\begin{center}
\includegraphics[width=\smwidth]{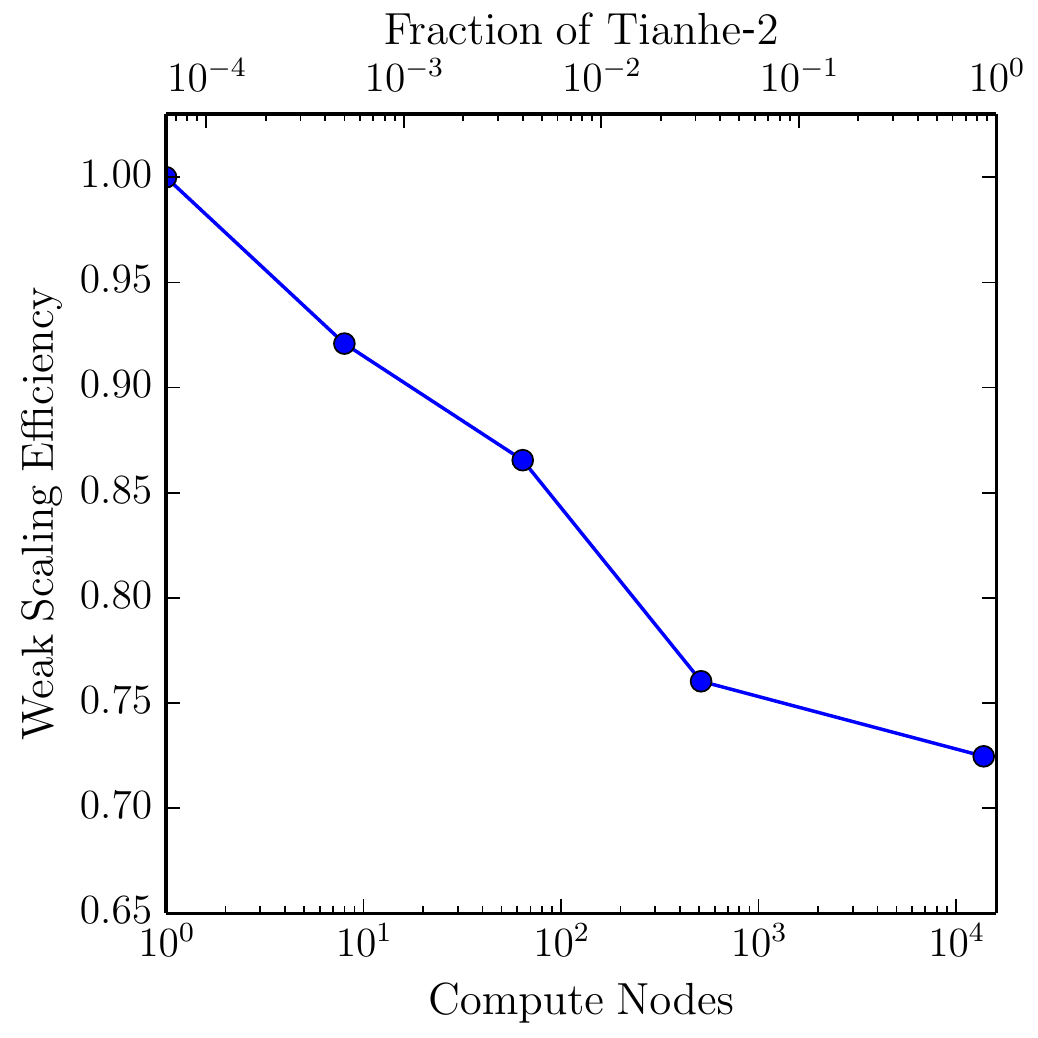}
\caption{Weak scaling efficiency of our cosmological neutrino simulations on Tianhe-2. In these trials,
each node initially contains $288^3$ CDM particles and $576^3$ neutrino particles in a periodic box of side
length $50\ h^{-1}{\rm Mpc}$. CDM is first evolved in isolation from $z = 100$ to $z = 5$ with neutrinos added
next. In each trial, we record the mean time step, $\bar{t}_{\rm step}$, of the first 400 steps after neutrinos are 
added and pivot the results against the single-node case.
We find a weak scaling efficiency of $72\%$ on 13,824 nodes ($86\%$ of the machine), 
corresponding to our production run, TianNu.}
\label{fig:scaling}
\end{center}
\end{figure}

\cpm\ is designed for weak scaling: if we hold the workload per process fixed, the wall-clock required
per time step should be roughly the same when we increase the number of processors in proportion to the problem size. 
Figure \ref{fig:scaling} shows the scaling efficiency as we weakly scale the single-node job described above. 
The weak scaling efficiency is defined to be $\bar{t}_0/\bar{t}_{\rm step}$ where $\bar{t}_{\rm step}$ is the average
wall-clock of each time step and $\bar{t}_0$ is the value of $\bar{t}_{\rm step}$ for the single-node case. Figure \ref{fig:scaling}
shows results for scaled versions of the single-node case on $n_{\rm mpi}^3 = 2^3$, $4^3$, $8^3$, and $24^3$ nodes of 
Tianhe-2 (recall that \cpm\ constrains the number of MPI tasks to be a perfect cube). Processor workload is held fixed by
increasing the total particle count and simulation volume in proportion to $n_{\rm mpi}^3$. In this way, the spatial resolution 
is constant and each simulation probes the same degree of non-linear physics. 

The code portrays a high degree of weak scaling efficiency all the way up to $24^3$ nodes where 
$\bar{t}_0/\bar{t}_{\rm step} = 72\%$. Each scaling trial contains the first 400 time steps from
the start of neutrino initialization at $z = 5$ (see below). Hence, each run contains a fair amount
of load imbalance in the CDM component as these particles have already evolved into the relatively
clustered regime. Neutrinos, on the other hand, are dispersed roughly equally throughout the
volume and remain this way for the duration of the simulation due to their thermal motion. This latter
fact tends to promote a relatively balanced workload across ranks. The high degree of scaling efficiency 
seen here is an encouraging result for the prospect of cosmological neutrino simulations. 


\subsection{TianNu Simulation}
\label{subsec:prodrun}

The rightmost point in Figure \ref{fig:scaling} corresponds to our production simulation, TianNu, containing 
$6912^3$ CDM particles and $13824^3$ neutrino particles in a periodic box of side length $1200\ h^{-1}{\rm Mpc}$. 
TianNu was run on 13,824 compute nodes with an equal number of MPI ranks and 24 OpenMP threads per rank.
This work is primarily focused on the computational aspects of preparing our code and running at scale on
Tianhe-2. However, since large-scale neutrino simulations have emerged only relatively recently in the field, 
we will provide here some further background information in regards to our neutrino implementation. 
We point the interested reader to \citet{inman/etal:2015} for more details about our initialization strategies and 
for a validation of our methods.

The most important difference in our strategy compared to pure CDM setups is the fact that neutrinos are initialized 
at a later time than CDM. This is done entirely from a practical standpoint and is sourced by safeguards 
in the code that limit the maximum distance a particle can travel in a single time step. At the time of CDM initialization, 
neutrinos in the mass range of interest are highly relativistic, meaning they move much faster than CDM and their
inclusion would slow down the simulation by an enormous amount. 
Our solution is to remove neutrinos in the early stages of the simulation while including their
contribution to the background expansion as a relativistic species. In the case of TianNu, CDM particles were initialized at
$z = 100$ and evolved in isolation to $z = 5$ for a total of 549 time steps. Neutrinos were then added into the
mixture\footnote{TianNu simulates neutrinos of mass $m_\nu = 0.05$ eV, which at $z = 5$ have a mean speed of 
$\langle v \rangle = 0.06c$ ($c$ is the speed of light), and thus are still modestly relativistic but now computationally
tractable to include in the simulation.} and the two
components evolved together until $z = 0$ for a total of 1918 time steps\footnote{See 
\url{http://cita.utoronto.ca/~haoran/thnu/movie.html} for an animation depicting the evolution 
in the two components followed by a flythrough of the simulation volume.}. The choice of the neutrino initialization redshift
is justified by the fact that, on the scales of interest, neutrinos are still within the linear regime of structure formation at 
$z = 5$, so their initial displacements and velocities are properly computed at that time. 
The majority of the simulation was spent in
the co-evolution stages ($z \leq 5$), accounting for $87\%$ of the total wall-clock time. TianNu consumed an effective
runtime of 52 hours though the actual amount of human time that elapsed from start to finish was much longer, due to
various obstacles addressed in the next subsection.

One important point worth noting is that unphysical particle coupling can occur in multi-species simulations when
unlike particle pairs are placed initially close together \citep[e.g.,][]{yoshida/etal:2003,angulo/etal:2013}.
This could have a potential impact on our simulation since some neutrinos may by chance be placed initially close to
CDM particles at $z = 5$; however, we expect this to be rare, given that CDM has evolved away from its initial lattice
configuration at $z = 5$. In addition, the large initial velocities of neutrinos should provide thermal support that protects
them against artificial coupling. In any case, our targeted science is focused on moderately large scales which should
remain robust to any such issues. To date, the analysis of TianNu resulted in two companion papers 
\citep{yu/etal:2016,inman/etal:2016} that present results on non-linear cosmological neutrino physics.

\subsection{Computational Challenges}
\label{subsec:techchallenges}

One of the difficulties associated with performing extreme-scale simulations are the unanticipated technical
setbacks that inevitably occur when a large fraction of the machine is used coherently at once. Some problems are
relatively easy to overcome with software changes while others involve external factors that require more careful
consideration. We present here the main technical challenges that we encountered as we scaled our simulation
to use $86\%$ of Tianhe-2. These problems are not specific to the cosmological problem at hand and are thus
relevant to the general field of scientific computing. 

Our first setback involved an apparent bug in the MPI library. We discovered this bug during an MPI\_Alltoall
call used to transpose data in the long-range PM force calculation. The bug was discovered early in our scaling tests
and seemed to be related to a single precision integer overflow within the MPI library. In our particular case, we were attempting
to send and receive $n$ elements of type MPI\_Real where $n > 2^{29}-1$. This resulted in a segmentation fault 
on Tianhe-2 as well as various other machines we tested using different MPI implementations (i.e., Intel MPI, Open MPI, 
and MVAPICH). We deduced that this error was caused by an internal calculation of the number of bytes being sent/received 
represented as a single precision integer. This was checked by changing the data type to MPI\_Double which failed for 
$n > 2^{28} - 1$. We were unable to remedy the problem by compiling in double precision and found this bug to be
a generic feature of all MPI routines, not just MPI\_Alltoall. Presently, the bug seems to be fixed, at least with Intel MPI v4, 
which we have explicitly checked at the time of writing.

At the time, our workaround was to manually replace the MPI\_Alltoall call with pairwise send/receive 
communications using MPI\_Send and MPI\_Recv. The message buffers were broken into pieces
such that no individual piece exceeded $2^{31}-1$ bytes. Initially, we attempted to use multiple non-blocking communications 
with MPI\_Isend and MPI\_Irecv, but found this created too much network strain,
leading to frequent system crashes when running TianNu. Blocking pairwise communication
resulted in much more stable data transmission, especially when using a large number of MPI ranks.
For this reason, we suspect our workaround is still better suited for
extreme-scale applications, despite the fact that the initial bug prompting us to abandon MPI\_Alltoall has
since been fixed. This, of course, depends on how internal communications are handled in MPI\_Alltoall, though
we now prefer the option of being able to explicitly make this choice in our code. In any event, we urge 
computational scientists to think carefully about communications when scaling their code, as this was one of our
main sources of grief.

The next setback we encountered involved runtime I/O. While we were only mainly interested in storing particle data
at the final output for analysis purposes, it was important to checkpoint frequently during runtime to ensure that progress
was maintained in the event of system crashes. Using our compressed data format, each TianNu checkpoint was 
roughly $25$ TB in total size, with this divided into several files for each MPI task. We experienced
a variety of issues while attempting to checkpoint in TianNu. The most common was a
system crash when one or more nodes failed during write, resulting in an incomplete checkpoint. In more
insidious cases, the checkpoint was seemingly successful and the code evolved forward, but further inspection showed
some files to be incompletely written. These were particularly difficult problems to debug since they required a large problem
size to occur frequently. 

Our attempt to circumvent these issues involved writing checkpoints to shared memory.
This was achieved by writing to the local dev/shm temporary filesystem on each node.
After all nodes completed this write, we used $100$ background processes to sequentially log into each of the
13,824 compute nodes and offload their checkpoint to the main filesystem while the simulation proceeded forward. 
This process is somewhat analogous to the operation of a burst buffer. 
Indeed, we found that using effectively only 100 processes to checkpoint put considerably
less pressure on the filesystem, resulting in much more reliable I/O.  This also significantly reduced the amount of time spent 
checkpointing since writing to shared memory is a relatively quick operation. Obviously, this is not a robust solution in all cases 
since it requires having sufficient memory to store a full checkpoint, and that the time between subsequent checkpoints is 
longer than the time required to offload to the filesystem.  In the end, we still encountered problems with this implementation, 
notably having trouble logging into some of the compute nodes at times. We are still working to perfect our
I/O implementation for future runs.

The final computational challenge we faced involved environmental factors. During our time on Tianhe-2, we had teams in both
China and Canada, which allowed for nearly continuous monitoring of TianNu. Interestingly, the team in Canada
had systematically more success in evolving the simulation forward. The reason was an increased level of system instability 
during Chinese daytime hours which lead to frequent system crashes that hindered progress. 
Tianhe-2 technical stuff speculated that this was due to increased ambient temperatures and more strain on
the electric grid during Chinese daytime hours. Accordingly, system crashes were somewhat alleviated during our second week on
Tianhe-2 when rainfall and cooler weather in Guangzhou seemed to be correlated with improved system stability. 
Regardless of the cause, external factors such as system instabilities are difficult to prepare for and have no clear solution. 
In our case, the only workaround was patience and persistence. 


\section{Conclusion}
\label{sec:conclusion}

Pushing the cosmological neutrino problem to extreme scales is a nontrivial process. A number of modifications to the
cosmological code \cpm\ were required to maximize performance on Tianhe-2. These included adopting a
two-dimensional pencil decomposition for parallel MPI Fourier transforms and implementing nested OpenMP parallelism to 
maximize multicore usage while maintaining memory flexibility. With these modifications, we achieved 72\% weak scaling 
efficiency on 13,824 nodes (331,776 CPU cores) of Tianhe-2. Our production simulation, named TianNu, consumed an effective
runtime of 52 hours on 86\% of the machine and pushed the cosmological neutrino problem forward by two orders of magnitude in
scale.

Data compression has become an increasingly important factor from the standpoint of runtime performance as well as data storage
and portability.  We have devised a novel method of data compression relevant for cosmological particle simulations. Our scheme provides
a compression factor of 4x and 2x in memory over single-precision positions and velocities, respectively, while maintaining sub-percent accuracy in density
and velocity power spectra for the vast majority of scales resolved in the simulation. This is true when compressing a single snapshot and
also when evolving particles forward in time from multiple previously compressed restarts.

Unanticipated challenges tend to emerge when scaling code to new limits. In our case, we encountered both software and hardware
problems. Two of these fall into the group of usual suspects: communication and I/O. An initial bug in the MPI library discovered early in our
scaling tests prompted us to break apart MPI\_Alltoall communications into buffers no larger than $2^{31}-1$ bytes. This turned out to be 
beneficial during the TianNu runtime since we could explicitly enforce blocking, pairwise communication; other communication strategies
were found to be unstable. Runtime I/O also proved finicky on extreme scales and required a makeshift burst-buffer approach where checkpoints
were written to RAM and offloaded to disk in the background. We are still working on a more robust solution for future applications. 
Finally, we experienced a high degree of time variability in machine performance due to local external factors. This is a considerably intractable
problem and cases like this should be kept in mind when preparing for extreme-scale simulations. 

The main challenge facing cosmological neutrino simulations can be summed up by two competing effects. 
On the one hand, large cosmological volumes are needed to acquire sufficient statistics to detect subtle neutrino 
effects on LSS. On the other hand, the thermal motion of neutrinos demands high particle number density to
suppress their shot noise on small scales. These factors can only be reconciled by running at extreme scale.
This work attempted to highlight the challenges associated with exploiting modern scientific computing to  
elucidate our understanding of neutrinos. 


\begin{acknowledgements}
Special Program for Applied Research on Super Computation of the NSFC-Guangdong Joint Fund (the second phase).
We thank Prof. Yifang Wang of IHEP for his great initial support of our project, and Prof. Xue-Feng Yuan for his kind and 
invaluable assistance during our time on Tianhe-2.
JDE, DI, and ULP gratefully acknowledge the support of the National Science and Engineering Research Council of Canada.
Work at Argonne National Laboratory was supported under U.S. Department of Energy contract DE-AC02-06CH11357.
HRY acknowledges General Financial Grant No.2015M570884 and Special Financial Grant No. 2016T90009 from the China Postdoctoral Science Foundation.
JHD acknowledges support from the European Commission under a Marie-Sk{\l}odwoska-Curie European Fellowship (EU project 656869).
XC acknowledges support from MoST 863 program 2012AA121701, NSFC grant 11373030, and CAS grant QYZDJ-SSW-SLH017.
TianNu was performed on the Tianhe-2 supercomputer at the National Super Computing Centre in Guangzhou, Sun Yat-Sen University.
This work was supported by the National Science Foundation of China (Grants No. 11573006, 11528306, 10473002, 11135009), 
the Ministry of Science and Technology National Basic Science program (project 973) under grant No. 2012CB821804, 
the Fundamental Research Funds for the Central Universities.
The simulations used in the error analysis of the data compression algorithm were performed on the GPC supercomputer at the
SciNet HPC Consortium.  SciNet is funded by: the Canada Foundation for Innovation under the auspices of Compute Canada; 
the Government of Ontario; Ontario Research Fund - Research Excellence; and the University of Toronto.
\end{acknowledgements}


\bibliographystyle{raa}
\newcommand{\noop}[1]{}

\end{document}